\documentclass[twoside]{article}
\usepackage[latin1]{inputenc}
\usepackage{t1enc}
\usepackage{a4}
\usepackage{tabularx}
\usepackage{epsf}

\def\bref{\vspace{4pt}\noindent\hangindent=10mm}

\begin{document}

\setcounter{figure}{0}
\setcounter{section}{0}
\setcounter{equation}{0}

\begin{center}
{\Large\bf
At the Bottom of the Main Sequence\\[0.2cm]
{\large Activity and magnetic fields beyond the threshold to complete convection}}\\[0.7cm]

Ansgar Reiners \\[0.17cm]
Georg-August-Universit\"at G\"ottingen, Institut f\"ur Astrophysik\\
Friedrich-Hund-Platz 1, 37077 G\"ottingen\\
\texttt{Ansgar.Reiners@phys.uni-goettingen.de}
\end{center}

\vspace{0.5cm}

\begin{abstract}
  \noindent{\it The bottom of the main sequence hosts objects with
    fundamentally different properties. At masses of about
    0.3\,M$_{\odot}$, stars become fully convective and at about
    0.08\,M$_{\odot}$ the hydrogen-burning main sequence ends; less
    massive objects are brown dwarfs. While stars and brown dwarfs
    experience very different evolutions, their inner structure has
    relatively little impact on the atmospheres. The generation of
    magnetic fields and activity is obviously connected to the
    threshold between partial and complete convection, because dynamo
    mechanisms involving a layer of shear like the solar
    $\alpha\Omega$-dynamo must cease. Hence a change in stellar
    activity can be expected there. Observations of stellar activity
    do not confirm a rapid break in activity at the convection
    boundary, but the fraction of active stars and rapid rotators is
    higher on the fully convective side. I summarize the current
    picture of stellar activity and magnetic field measurements at the
    bottom of the main sequence and present recent results on
    rotational braking beyond.}
\end{abstract}

\section{Introduction}

Low-mass stars and sub-stellar objects are fascinating astrophysical
laboratories affected by various physical processes. The inner
structure of low-mass stars undergoes two very interesting phase
transitions that seriously influence their structure and evolution.
The first transition occurs around 0.35\,M$_\odot$ (spectral type
$\approx$~M3.5, main-sequence effective temperature
$\approx$~3500\,K); stars less massive are completely convective while
more massive stars possess an inner radiative zone similar to the Sun.
Because the solar (interface) dynamo is closely related to the shear
layer between the two regions (the tachocline, Ossendrijver, 2003),
cooler stars lacking a tachocline cannot maintain such a Sun-like
interface dynamo.  The second important threshold occurs around
0.08\,M$_\odot$, the dividing line between stars and brown dwarfs.
Less massive objects do not produce enough heat in their interior to
ignite stable hydrogen fusion (e.g., Chabrier \& Baraffe, 2000).  They
keep cooling while contracting which makes them wander through later
and later spectral classes getting fainter and fainter.

The spectral appearance of low mass stars is governed by two
atmospheric phase transitions. As temperature drops in the atmospheres
of cool stars, molecules start to form around effective temperatures
of some 4000\,K. At about 2500\,K, dust grains rain out making the
(sub)stellar atmospheres even more complex, and the growing neutrality
of the atmosphere entirely changes its physical properties.

Understanding the physics of objects at and beyond the bottom of the
main sequence is particularly challenging because these objects are
relatively difficult to observe. Although many are close by they are
so dim that large telescopes are necessary to uncover their secrets.
Most of our knowledge about the physics of stars comes from
spectroscopy.  Spectra of ultra-cool objects, however, are dominated
by molecular absorption, and isolated spectral lines suitable for
detailed investigation -- a standard in hotter stars -- are hardly
available.

The physical richness of the bottom at the main sequence and the
observational problems are nicely summarized in the review by Liebert
\& Probst (1987) several years before the first brown dwarf was even
discovered. In the 20 years since then large aperture telescopes and
much improved observing facilities, sensitive to infrared wavelengths,
revealed a huge amount of information on low-mass objects. Not only
have we learned that brown dwarfs are actually reality, we have also
seen that their formation mechanism as well as their complex
atmospheres are indeed as difficult to understand as expected by
Liebert \& Probst (1987). The history of brown dwarf observations and
our knowledge about them is reviewed for example in the articles by
Basri (2000), Kirkpatrick (2005), and Chabrier et al. (2005). Here, I
will not try to give a comprehensive update on the bottom of the main
sequence. I will rather focus on one point: Does the physically
important threshold to complete convection have visible effects on
low-mass stars?

\subsection{Convection in the HR-diagram}

\begin{figure}[t]
  \epsfxsize=\textwidth \epsfbox{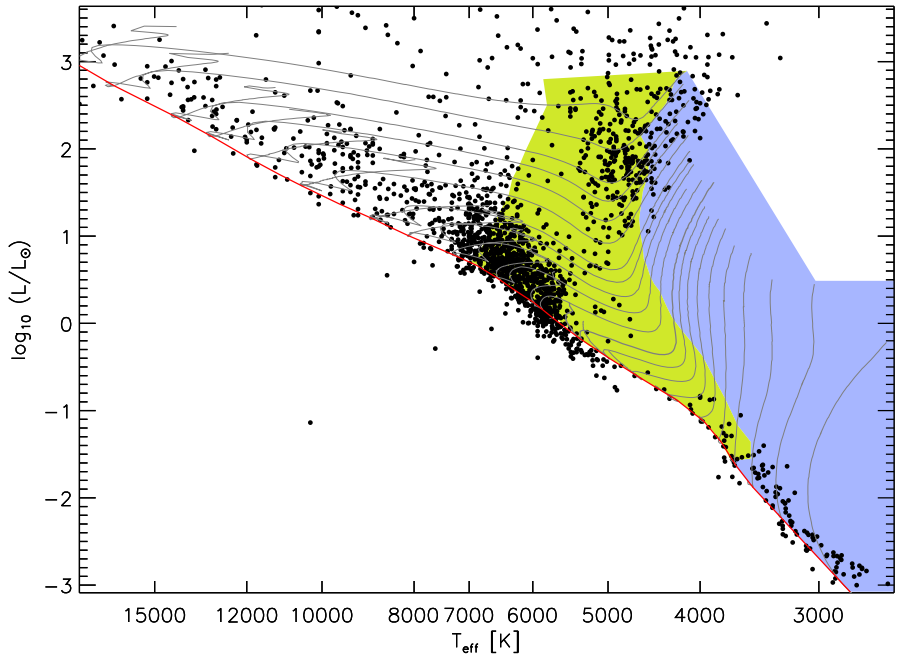}
  \caption{\label{fig:HRdiagram}HR-diagram with evolutionary tracks
    from Siess et al.  (2000). Regions where stars have outer
    convective shells are indicated in 
green    
    color, stars occupying the 
blue
    region are completely convective. The ZAMS is shown with a 
red 
    line.}
\end{figure}

The presence of convection governs a star's ability to maintain
significant magnetic fields that are the reason for the wide range of
phenomena summarized as magnetic activity. In the absence of an
(outer) convection zone, no dynamo can generate magnetic flux anywhere
close to the surface so that fossil fields are the most plausible
candidates for the strong magnetic fields observed in hot stars as for
example Ap-stars.

In Fig.\,\ref{fig:HRdiagram}, I show an HR-diagram with pre-main
sequence evolutionary tracks taken from Siess et al. (2000).  Stars
plotted in this diagram are mainly from the Hipparcos catalogue (ESA,
1997) with temperatures from Taylor (1995), Cayrel de Strobel (1997),
or converted from $uvby\beta$-colors from Hauck \& Mermilliod (1998).
Additional low-mass stars are taken from Bessel (1991) and Leggett
(1992).  For the interesting question of stellar dynamos and their
interaction with rotation, the question whether stars have convective
envelopes or are fully convective is essential. I have indicated in
Fig.\,\ref{fig:HRdiagram} regions in the HR-diagram where stars have
outer convective envelopes
(green) 
and where they are completely convective 
(blue)
during pre-main sequence evolution and on the main sequence (Siess et
al.  2000). The temperature region at which stars develop a convective
shell is around 5800--6500\,K, this is governed by the ionization of
hydrogen. On the main sequence, the threshold to complete convection
happens around $T_{\rm eff} = 3500$\,K but at much higher temperatures
in younger stars. Essential information on the nature of Sun-like and
fully convective dynamos can be expected by investigating activity
close to these two thresholds.

\subsection{An extended HR-diagram}

\begin{figure}[t]
  \epsfxsize=\textwidth \epsfbox{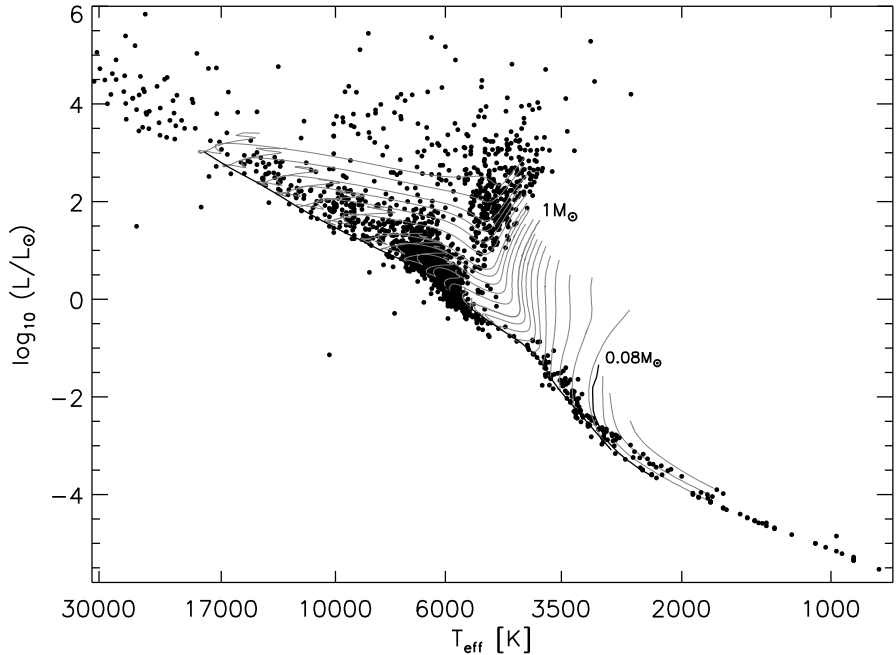}
  \caption{\label{fig:extHRdiagram} Extended HR-diagram including very
    low-mass objects. Evolutionary tracks from Baraffe et al. (1998
    and 2002) are added. The evolution of an 0.08\,M$_{\cdot}$ star is
    shown with a black line.}
\end{figure}

During the last years, temperature calibrations of low-mass objects of
spectral class L and T became available (Dahn et al. 2002; Golimowski
et al. 2004). We can use this new information to extend the HR-diagram
towards the coolest known objects. Fig.\,\ref{fig:extHRdiagram} shows
an HR-diagram covering temperatures well below 1000\,K.  Additional
evolutionary tracks from Baraffe et al. (1998 and 2002) are shown. The
temperatures from Dahn et al. (2002) were shifted by 400\,K to the hot
side in order to achieve consistency (see Golimowski et al 2004). The
evolutionary track for a 0.08\,M$_{\odot}$ object is plotted as a
thick line.  It demonstrates a crucial feature in the analysis of
brown dwarfs; because hydrogen burning cannot stabilize them, brown
dwarfs become cooler and dimmer during their entire lifetime. They
never maintain a constant temperature over a (cosmologically) long
period. In other words, a certain temperature is not indicative of a
certain mass in very low mass objects (VLMs), and because the
transition from stars to brown dwarfs is not abrupt, objects of
temperatures around 2500 -- 3100\,K can either be old stars or young
brown dwarfs.

The luminosity of main sequence stars hotter than about 4000\,K is
declining smoothly. In the brown dwarf regime, i.e. at temperatures
lower than $T_{\rm eff} = 2500$\,K, the ``main sequence'' also follows
a well-defined but shallower slope. Here it coincides with the line of
constant radius, $R \approx 0.1$\,R$_{\odot}$; all objects in this
regime have approximately the same radius because of electron
degeneracy. Between these two regimes, around $T_{\rm eff} = 3500$\,K,
a step in luminosity appears, which was mentioned, e.g., by Hawley et
al.  (1996). It occurs close to the temperature of H$_2$-association
(Copeland et al. 1970), but theoretical expectations including this
effect differ from the observations (e.g., D'Antona \& Mazzitelli
1996; Baraffe \& Chabrier 1996). The step around $T_{\rm eff} =
3500$\,K shows a remarkable coincidence with the onset of complete
convection (cp Fig.\,\ref{fig:HRdiagram}). Clemens et al.  (1998)
discuss this point and its possible influence on the period
distributions of binaries.

\section{Rotation and magnetic activity in very low-mass objects}

The rotation-activity relation has become one of the more solid
concepts in stellar astrophysics. In sun-like stars, i.e. stars with
an outer convective shell, virtually all tracers of chromospheric and
coronal activity correlate with rotation (e.g., Ayres \& Linsky, 1980;
Noyes et al., 1984; Simon, 2001; Pizzolato, 2003; and references
therein). Activity is connected to stellar rotation in the sense that
the more rapidly a star rotates, the more emission in tracers of
activity is observed until a saturation level is reached beyond which
emission does not become stronger anymore (maybe it becomes even less
at very high rotation in the ``supersaturation'' regime; e.g.
Randich, 1998). The question whether and how the rotation-activity
relation is universal for all types of stars may be subject of debate
(e.g., Basri, 1986), but it seems clear that the rotation period is
the most important individual parameter for the generation of magnetic
activity.  The general picture is that of a solar-like dynamo mainly
of $\alpha-\Omega$ type. The efficiency of magnetic field generation
through this dynamo depends on rotation rate (or Rossby number, Durney
\& Latour, 1978); more magnetic flux is produced at rapid rotation.
Magnetic flux is believed to generate all phenomena of stellar
activity in analogy to the solar case.

In order to understand stellar dynamo processes it is always
instructive to investigate stars with very different physical
conditions for a dynamo to work. A key area is where stars become
completely convective. The ``classical'' $\alpha-\Omega$ dynamo,
responsible at least for the cyclic part of solar activity, is
believed to be situated at the tachocline between the radiative core
and the convective envelope. This dynamo mechanism must cease towards
lower mass objects.

\subsection{Activity}
\label{sect:activity}

Stellar activity is quantified through the strength of emission
observed either in broad wavelength regions (e.g., X-ray) or in
specific spectral lines.  Which tracer is most suitable for the
investigation of stellar activity is a question of the emission
produced by the star and a question of the contrast to the spectral
energy distribution of the photosphere.

X-rays are well suited indicators in a broad range of stars, because
the blackbody emission from the star is virtually free of X-ray
emission (i.e. the contrast is very high). X-ray observatories,
however, are comparably small (and few in number which reduces the
available observing time) so that the detection threshold for X-ray
emission is relatively high. It has been found that the ratio of X-ray
luminosity $L_{\rm X}$ to bolometric luminosity $L_{\rm bol}$ even in
the most active stars rarely exceeds a ratio of 10$^{-3}$ (e.g.
Pizzolato et al., 2003, and references therein). This means that X-ray
emission from faint (and small) VLMs with low $L_{\rm bol}$ is more
difficult to probe than from hotter ones. X-ray measurements in field
surveys are available for objects as late as mid-M, only sparse
information is available for cooler ones.

In the Sun, X-rays are produced in the corona and we usually assume
similar emission processes in other stars. In sun-like stars, coronal
emission is closely connected to chromospheric emission. The deep
absorption cores of the Ca\textsc{II} H\&K lines, Mg\textsc{II} h\&k
lines, and other UV-lines are good regions to look for chromospheric
emission. In very cool objects, however, H$\alpha$ is the line of
choice partly because the contrast to the photosphere is much weaker
than in hotter stars; it is rather strong and very easy to observe.

The vast majority of X-ray detections in low mass stars comes from the
ROSAT mission (see for example Schmitt, 1995, and references therein).
Unfortunately, the sensitivity of ROSAT did not allow to measure X-ray
emission in many objects of mid-M spectral type or later. A number of
measurements in selected active ultra-cool objects were carried out
with Chandra and XMM (e.g., Stelzer, 2004, and references therein;
Robrade \& Schmitt, 2005). Normalized X-ray emission between
$\log{L_{\rm X}/L_{\rm bol}} = -2$ and $-4$ was detected during flares
in some objects. The overall (quiet) emission level, however, is
decreasing (see Fig.\,4 in Stelzer et al., 2004) and mostly below the
detection level. Emission observed in the H$\alpha$ line is an
indicator available for a much larger number of VLMs. Flares in very
low mass objects reach a comparable level in normalized H$\alpha$
emission as observed in some X-ray observations.  For a first
comparison, it is probably not too wrong to assume that observations
of H$\alpha$ emission probes similar mechanisms as X-ray observations
do.

\begin{figure}[t]
  \epsfxsize=\textwidth \epsfbox{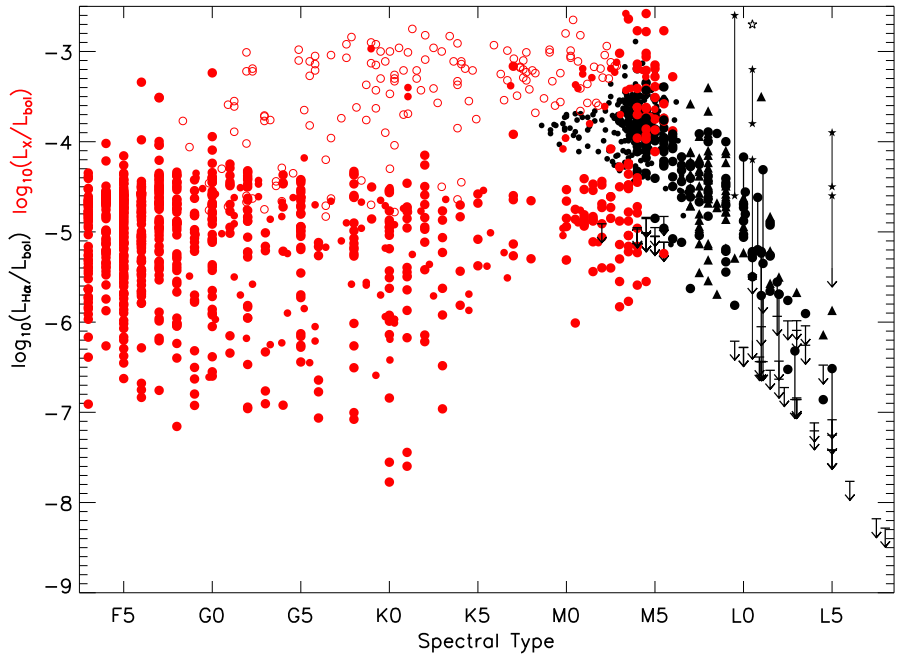}
  \caption{\label{fig:activity} Normalized activity vs. spectral type.
    X-ray and H$\alpha$ emission are shown on the same scale.  Lines
    connect observations of identical objects. X-ray measurements 
(red)
    are from Voges et al., 1999 (field stars, filled circles) with spectral
    types from the Hipparcos catalogue (ESA, 1997). Open circles are
    cluster stars from Pizzolato et al.  2003, field stars from the
    same publication are plotted as filled circles.  H$\alpha$
    (black): small circles from Reid et al., 1995 and Hawley et al.,
    1996; large circles from Mohanty \& Basri, 2003, Reiners \& Basri,
    2007, Reiners \& Basri, in prep.; triangles from Schmidt et al.,
    2007; stars from Burgasser et al., 2002 and Liebert et al., 2003.}
\end{figure}

So far, in order to draw a comprehensive picture of stellar activity
from early F stars to brown dwarfs, we have to combine the results
from different tracers.  Reiners \& Basri (2007) showed that X-ray and
H$\alpha$ measurements among M-stars exhibit a rough correspondence,
i.e. for a first estimate we may assume $L_{\rm X}/L_{\rm bol} \approx
L_{\rm H\alpha} / L_{\rm bol}$. In hotter stars, however, Takalo \&
Nousek (1988) found an offset of about 1.8\,dex between these two
indicators (in hotter stars there is also a close correspondence
between Ca\textsc{II} H\&K and X-ray emission as found by Sterzik \&
Schmitt, 1997). Hawley et al. (1996) found a constant offset between
$L_{\rm X}$ and $L_{\rm H\alpha}$ of only about half a dex in
late-type stars. A census of currently available X-ray and H$\alpha$
measurements is plotted in Fig.\,\ref{fig:activity}.  Although one
has to keep in mind that the two tracers, plotted in 
red 
and black
, do not match each other directly (see above), the error introduced
is probably not larger than $\sim$0.5\,dex at spectral types where
more H$\alpha$ measurements become available and X-ray measurements
become rare. In any case, the correspondence between $L_{\rm X}$ and
$L_{\rm H\alpha}$ is probably not too bad, and their direct comparison
is very instructive.

The current picture of stellar activity may be summarized with the
help of Fig.\,\ref{fig:activity}: Among the sun-like stars earlier
than M0, a clear age-rotation-activity dependence exists. Young
(cluster) stars show normalized X-ray luminosities $L_{\rm X}/L_{\rm
  bol}$ roughly between $-3$ and $-4$, they are rapid rotators in the
saturated part of the rotation-activity connection (see Patten \&
Simon, 1996; Pizzolato et al., 2003). The older field stars have
normalized X-ray luminosities an order of magnitude lower than their
younger predecessors, i.e., $L_{\rm X}/L_{\rm bol} \le 10^{-4}$ in the
field. The level of activity can be quite low and probably all
sun-like stars show some sort of activity that could be measured if
the sensitivity was high enough. Activity scales with rotation period
or Rossby number, which is rotation period divided by the convective
overturn time.  Among F--K stars, the convective overturn time does
not change dramatically so that relations in rotation period or Rossby
number are not too different.

\begin{figure}[t]
  \center
  \epsfxsize=.9\textwidth \epsfbox{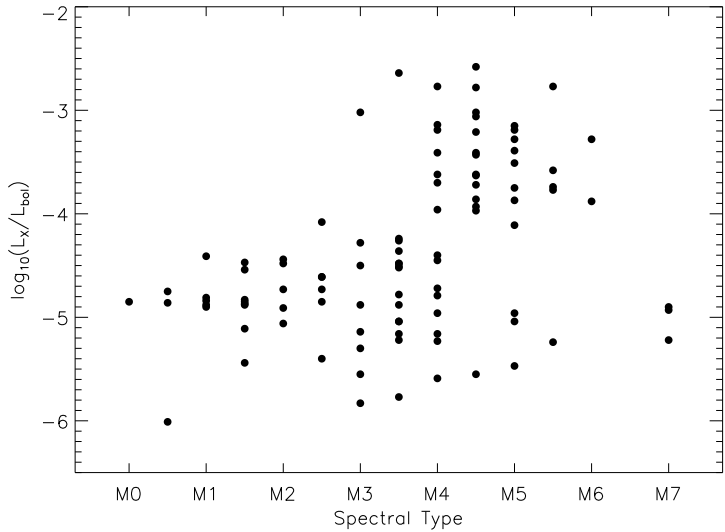}
  \caption{\label{fig:Delfosse}Normalized X-ray activity as a function
    of spectral class for M-type stars. Data are taken from Delfosse
    et al., 1998.}
\end{figure}

A change in normalized X-ray activity among the field stars is visible
in Fig.\,\ref{fig:activity} between spectral types M0 and M5 (both in
X-ray and in H$\alpha$).  Around spectral type M5, normalized X-ray
luminosity ramps up and some mid-M stars exhibit values as high as
$\log{L_{\rm X}/L_{\rm bol}} > -3$. Stars in the spectral region
M0--M7 from the volume-limited sample of Delfosse et al. (1998) are
shown separately in Fig.\,\ref{fig:Delfosse}. This plot suggests a
rise in maximum normalized X-ray luminosity between spectral types M2
and M4 (a similar behavior is visible in the H$\alpha$ data in Fig.\,2
of Hawley et al., 1996). This spectral range coincides with the mass
range where stars become completely convective. The rise in X-ray
activity is probably not the consequence of an entirely different
process of magnetic field generation. If that was the case, it would
imply that stars in the completely convective regime are even
\emph{more} efficiently generating magnetic fields, which is difficult
to believe. As shown in Section\,\ref{sect:rotation}, the reason for
the rise in activity is more likely weaker rotational braking in
completely convective stars.  Fields stars beyond spectral type M4 are
generally more rapidly rotating, and there is ample evidence that the
rotation activity connection still applies in fully convective mid-M
stars (Mohanty \& Basri, 2003; Reiners \& Basri, 2007). Thus, more
rapid rotation leads to higher X-ray emission. I will further discuss
what happens at the threshold to complete convection in
Section\,\ref{sect:convective}.

From the growing number of H$\alpha$ observations in VLMs, it is clear
that around spectral type M9, normalized H$\alpha$ emission gradually
decreases with spectral type, i.e. with temperature.  This effect is
probably not directly related to a lack of magnetic flux, but can
rather be explained by the growing neutrality of the cold atmospheres
(Meyer \& Meyer-Hofmeister, 1999; Fleming et al., 2000; Mohanty et
al., 2002).

Another way to search for stellar activity is looking for radio
emission. G\"udel \& Benz (1993) showed that in stars between spectral
class F and early-M, X-ray and radio emission are intimately related.
A probable explanation is gyrosynchrotron emission of mildly
relativistic electrons. Plasma heating and particle acceleration
probably occur in the same process. Berger (2006) searched for radio
emission in a sample of low mass objects. Although the sensitivity in
terms of normalized radio luminosity $\log{L_{\rm rad}/L_{\rm bol}}$
would not allow the detection of radio emission according to the
scaling found by G\"udel \& Benz (1993), Berger (2006) finds much
stronger radio emission in some very low mass objects. Hallinan et al.
(2006) argue that the strong and modulating radio emission observed in
some very low mass objects can be explained by coherent radio emission
(not incoherent emission). The magnetic fields they derive for very
low mass objects are on the order of kilo-Gauss which is consistent
with the observations by Reiners \& Basri (2007) and higher than the
estimates from incoherent emission (see Berger, 2006, 2007).

\subsection{Rotation}
\label{sect:rotation}

Rotation is generally believed to be intimately connected to stellar
activity through a dynamo process depending on rotation rate. The more
rapidly a star rotates the more magnetic flux is produced leading to
enhanced activity. At a certain rotation rate, activity reaches a
saturation level beyond which activity remains at the same level
independent of the rotation rate. At very high velocities, activity
may even ``supersaturate'', i.e. fall below the saturation limit again
(Patten \& Simon, 1996; Randich, 1998). The level of the saturation
velocity is thought to scale with convective overturn time, i.e.
saturation sets in at constant Rossby number. Combined with the
smaller radius of lower mass stars, this results in a drastic
divergence of saturation velocities at the stellar surface.  In early
G-stars, this velocity is on the order of 25--30\,km\,s$^{-1}$ while
in mid-M dwarfs it is less than 5\,km\,s$^{-1}$ (e.g., Reiners, 2007).
Such small rotation velocities are difficult to measure particularly
in M dwarfs so that the unsaturated part of the rotation activity
connection is not very well sampled (but it still holds, see Reiners,
2007).

Rotation rates or surface velocities can be measured through
rotational line broadening. This method, however, is only sensitive to
the projected rotation velocity $v\,\sin{i}$ and is limited by the
resolving power of the spectrograph (typically to $v\,\sin{i} \approx
3$\,km\,s$^{-1}$ in M dwarfs). Rotation periods can directly be
detected if stars have (temperature) spots stable enough to persist
longer than the rotation period so that the brightness modulation
induced can be followed for some rotations. In very cool stars, this
is a difficult task because the temperature contrast is probably very
low so that the amplitude of the brightness modulation is very small.
Detecting rotation periods in old M-stars turned out to be a very
frustrating business with only a few successful attempts so far (e.g.,
Pettersen, 1983; Torres et al., 1983; Benedict et al., 1998; Kiraga \&
Stepie\'n, 2007). One explanation for this is that lifetimes of spots
in very low mass stars are probably short while rotation periods are
relatively long.  Such spots will not show the same configuration
after one full rotation causing non-periodic brightness variations.
Satellite missions like \emph{COROT} and \emph{Kepler} will provide a
fresh look into rotation periods of low mass stars. Among the younger
objects, rotation periods from cluster measurements were much more
successfull (see e.g. Scholz \& Eisl\"offel, 2004; 2007; Herbst et
al., 2007 and references therein), and the rotational evolution at the
first few hundred million years is much better established than at
later phases.

Very active M-stars, dMe stars, show strong brightness variations on
very short timescales. Such ``flares'' are observed in outbursts of
emission lasting minutes to hours. Very active dMe stars show flaring
activity every some hours. These stars are generally rapid rotators
with rotation velocities on the order of a few km\,s$^{-1}$, and
rotational line broadening can be relatively easy to measure. They
occupy the saturated part of the rotation-activity connection.

\begin{figure}[t]
  \epsfxsize=\textwidth \epsfbox{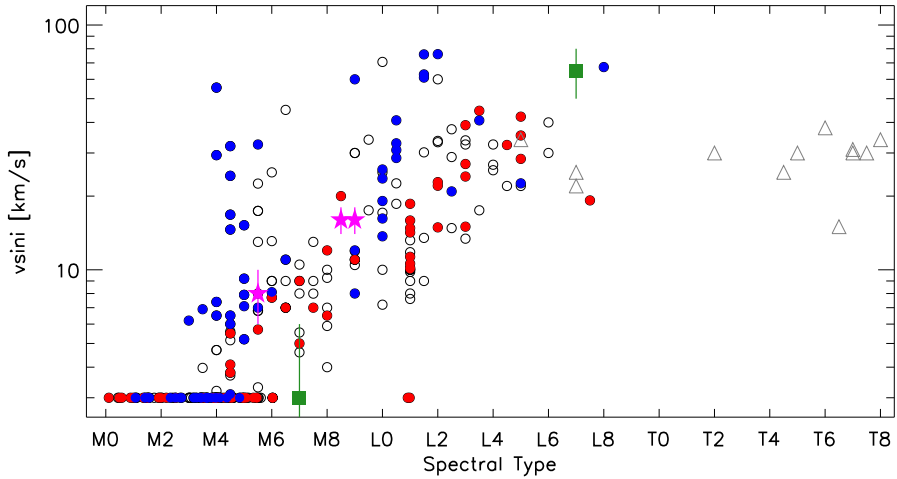}
  \caption{\label{fig:rotation}Projected rotation velocities in very
    low mass objects. Circles: Field M- and L-dwarfs from Reiners \&
    Basri, 2007; Mohanty \& Basri, 2003; Delfosse et al., 1998;
    Reiners \& Basri, in prep. Filled 
blue 
    and 
red 
    circles show objects that are probably young and old,
    respectively.  Open symbols are objects with no age information
    available.
Green squares: 
    Subdwarfs from Reiners \& Basri, 2006a.  
Magenta stars:
    The three components of LHS~1070 from Reiners et al., 2007.
    Open triangles: Field L- and T-dwarfs from Zapatero Osorio et al.,
    2006.}
\end{figure}

Less active M dwarfs are rotating very slowly. Their surface rotation
is on the order of one km\,s\,$^{-1}$ and hence very difficult to
measure spectroscopically (Reiners, 2007), because a resolving power
of more than $10^5$ is required. Rotation periods are on the order of
days to weeks, which in the absence of strong brightness modulation
and probably short lived spots (if any) is even more difficult to
detect.  For the reasons given above, measurements of rotation rates
in M dwarfs are almost exclusively available for ``rapid'' rotators,
not much is known about the distribution of rotation velocities below
$v\,\sin{i} \approx 3$\,km\,s$^{-1}$.

In Fig.\,\ref{fig:rotation}, I show a collection of measured projected
rotation velocities, $v\,\sin{i}$, in field objects (for references
see caption to Fig.\,\ref{fig:rotation}). The distribution of
$v\,\sin{i}$ among VLMs shows two remarkable features: (1) A sudden
rise in rotation velocities at spectral type M3.5/M4; and (2) a rising
lower envelope of minimum rotation velocities in the mid-M and L dwarf
regime. I discuss the two features in the following sections.

\subsubsection{The M dwarf spin-down puzzle}

Delfosse et al. (1998) obtained projected rotation velocities in a
volume-limited sample of roughly 100 field M dwarfs. Although they
come to the conclusion that ``the present data show no obvious feature
in the rotational velocity distribution at this type [around spectral
type M3], or elsewhere within the M0--M6 range'', their Fig.\,3 shows
quite a remarkable feature: The old disk and Halo population exhibit
very low rotation velocities (with possibly a gradual increase at
spectral types later than M5). With only one exception (at spectral
type M8), all old objects including the latest spectral types have
$v\,\sin{i} < 10$\,km\,s$^{-1}$. On the other hand, the young disk
population of field M dwarfs exhibits a break in the distribution of
rotation velocities. All young population members earlier than
spectral type M3 show low rotation rates (projected rotation
velocities on the order of the detection limit), but several rapid
rotators with $v\,\sin{i} > 20$\,km\,s$^{-1}$ are found at spectral
types M4 and later. Delfosse et al. conclude that the spin-down
timescale ``is of the order of a few Gyrs at spectral type M3--M4, and
of the order of 10 Gyr at spectral type M6''.

The sample of Delfosse et al. (1998) is included in
Fig.\,\ref{fig:rotation}. The break in the $v\,\sin{i}$ distribution
around spectral type M3.5 is clearly visible: In this compilation, all
stars earlier than spectral type M3 rotate slower than the detection
limits (usually around $v\,\sin{i} = 3$\,km\,s$^{-1}$).  Stars of
spectral type M4 or later exhibit rotation speeds up to
60\,km\,s$^{-1}$ and more in some exceptional cases. In general, the
rapid rotators belong to the young population 
(blue circles), 
i.e., rotational braking is still functioning at mid-M spectral
class, but it is obviously much less efficient than in hotter stars.

The ``standard'' theory of angular momentum evolution assumes that
stars accelerate during the first million years because of
gravitational contraction. Once on the main sequence, angular momentum
loss by a magneto-thermal wind brakes the star. Chaboyer et al.
(1995) and Sills et al. (2000) provide a prescription for angular
momentum loss during stellar evolution. Angular momentum loss is
proportional to some power of the angular velocity $\omega$ (Mestel,
1984; Kawaler, 1988). The power law itself depends on the magnetic
field geometry with very strong braking in the presence of a radial
field and lower braking if the field is dipolar.  Skumanich-type
rotational braking with $v \propto t^{-1/2}$ can be achieved by a
magnetic topology between these two cases. In this model, magnetic
braking is assumed to be proportional to some power of the angular
velocity $\omega$ as long as $\omega$ is small. At a critical angular
velocity, $\omega_{\rm crit}$, the relation between angular momentum
loss and $\omega$ changes (see, e.g., Sills et al., 2000). One choice
of the scaling of $\omega_{\rm crit}$ is given in Krishnamurti et al.
(1997).  They assume that $\omega_{\rm crit}$ is inversely
proportional to the convective overturn timescale (implying that
saturation sets in at constant Rossby number).

The model of angular momentum evolution is very successful in sun-like
stars of very different ages and a variety of spectral classes (e.g.,
Barnes, 2007). The main parameters governing rotational braking in
this model are the strength of the magnetic field and its geometry.
If a similar braking law applies to completely convective stars -- and
there is no reason to believe the opposite -- one of the two
parameters must dramatically change at least in the young population
of fully convective stars. As mentioned above, there is no reason to
believe that the magnetic field strength has a discontinuity at
spectral type M3.5 (maybe later it does), and I will discuss direct
observations of magnetic fields in completely convective objects in
Section\,\ref{sect:fields}.

\subsubsection{L-dwarf rotation and the age of the galaxy}

The second striking feature of the velocity distribution
(Fig.\,\ref{fig:rotation}) is the rising lower envelope of rotation
velocities at spectral types M7--L8. At spectral type M7, some stars
still show very low rotation velocities on the order of the detection
limit (about 3\,km\,s$^{-1}$). In cooler objects, however, the lowest
rotation velocities grow to about 10\,km\,s$^{-1}$ around spectral
type L2 up to the order of several ten km\,s$^{-1}$ at spectral type
L6--L8. In the L-dwarf data shown in Fig.\,\ref{fig:rotation} (mostly
from Mohanty \& Basri, 2003; Reiners \& Basri, in prep.), this lower
envelope is rather well defined with only two objects falling below a
virtual line from M7 (zero rotation) up to L8 (about
40\,km\,s$^{-1}$).  The two ``outliers'' at spectral types L1 and L7.5
may be seen under small inclination angles. Zapatero Osorio et al.
(2006) measured rotation velocities in a sample of brown dwarfs
finding two L7 dwarfs at comparably low rotation velocities.
Furthermore, they measured rotation velocities between 15 and
30\,km\,s$^{-1}$ in a couple of T dwarfs. To what extent systematic
effects due to the different techniques used affect the results should
not be discussed here but has to be kept in mind for interpretating
these results.  Nevertheless, there is agreement that rotation
velocities in VLMs are much higher than in hotter stars, which implies
that rotational braking is probably much weaker.

Is the rise of minimum rotational velocities with later spectral type
indeed a physical effect, or could it be due to an observational bias?
Brown dwarfs do not establish a stable configuration like stars do.
Young brown dwarfs are much brighter than old brown dwarfs so that a
brightness limited survey (as here) in general favors young brown
dwarfs. The observed rise in minimum rotational velocities could mean
that, for example, at spectral type L6 we are only observing young
(bright) objects that are still not efficiently braked, while at
earlier spectral classes we can already reach older objects that
suffered braking for a much longer time. In this picture a lack of
slowly rotating late-L and T dwarfs would simply mean that such old
objects are not contained in our sample because they are too faint.

The fact that brown dwarfs cool during their entire lifetime makes the
interpretation of the brown dwarf rotational velocity distribution
much more complex than the stellar one. There, evolutionary tracks in
the $v\,\sin{i}$ / spectral-type diagram essentially are vertical
lines on which the stars rise and fall during phases of acceleration
and braking. Brown dwarfs, on the other hand, migrate through the
spectral classes, they start somewhere between mid- and late-M
spectral class and eventually cool down to T-type or even later
spectral class. Thus, the distribution of rotation velocities with
spectral class among brown dwarfs is a function of rotational
evolution, coolings tracks, and of formation rates.

In Fig.\,\ref{fig:rotation}, age information is included for a number
of brown dwarfs.  Although sparse, the age information strongly
suggests that the rise of the lower envelope of minimum rotation
velocities is not an observational bias but rather a consequence of
rotational braking. If the brightness limit was the reason for the
observed distribution of rotation velocities, we would expect to see
only young objects among the mid- to late-L spectral classes (those
would actually be of planetary mass).  Old objects would only be
visible at earlier spectral types.  Instead, the age information of
our sample clearly shows that the entire lower envelope of rotation
velocities is occupied mainly by objects of the old disk or Halo
populations, while stars rotating more rapidly are generally younger
at each spectral type. This indicates that tracks of rotational
evolution go from the upper left to the lower right in
Fig.\,\ref{fig:rotation}. Thus, brown dwarfs are indeed being braked
during their evolution.

This interpretation of Fig.\,\ref{fig:rotation} is also supported by
the five individual objects contained. The three objects plotted as
stars are the three members of LHS\,1070 (GJ\,2005, Reiners et al.,
2007b). LHS~1070\,B and C have very similar spectral types (around M9)
while the A-component is a little more massive (M5.5).  There are good
arguments that the three components are of same age, and that they are
seen under comparable inclination angles. The two cooler objects show
similar projected rotation velocities of $v\,\sin{i} \approx
16$\,km\,s$^{-1}$ while the earlier one is rotating at half that pace.
This result also supports the idea that isochrones in the
rotation/spectral-type diagram run from the lower left to the upper
right. A possible explanation for this behavior is mass-dependent
rotational braking. The two objects plotted as filled squares in
Fig.\,\ref{fig:rotation} are subdwarfs that are probably among the
oldest objects in our galaxy (Reiners \& Basri, 2006a). While the
earlier subdwarf at spectral type sdM7 exhibits very low rotation, the
sdL7 is rotating at the remarkably high speed of $v\,\sin{i} \approx
65$\,km\,s$^{-1}$.  After the long lifetime of probably several Gyrs,
the sub-L dwarf has not significantly slowed down, which means that
rotational braking in this object must be virtually non-existing.

It is important to realize that, if the lower envelope of minimum
rotational velocities is occupied by the oldest brown dwarfs, these
did not have enough time to decelerate any further. Thus, the lower
envelope of rotational velocities is directly connected to the age of
the galaxy, and it should extend to lower $v\,\sin{i}$ in older
populations.

Zapatero Osorio et al. (2006) show the rotational evolution of brown
dwarfs through the spectral classes M, L, and T in the absence of
braking. Starting at velocities of $v\,\sin{i} = 10$\,km\,s$^{-1}$,
their test objects spin up due to gravitational contraction to the
final rotation speed of some ten km\,s$^{-1}$. This picture is in
qualitative agreement with the overall distribution of rotation
velocities among the latest brown (T-)dwarfs, but not with the
rotation/age distribution of L dwarfs.  Nevertheless, the simulations
of Zapatero Osorio et al. show that the rotation of T dwarfs may be
explained by gravitational contraction in the absence of any braking.
The initial rotation velocities between the L and T dwarfs populations
must then be different, because this model cannot explain rotation
velocities as high as observed in the mid-M/early-L type objects.

\subsection{Magnetic Fields}
\label{sect:fields}

In order to understand the physical processes behind stellar magnetic
activity, it is obviously desirable to directly measure magnetic
fields. Unfortunately, the direct measurement is much more difficult
than measuring most of the indirect activity tracers. The direct
measurement of stellar magnetic fields usually means to measure Zeeman
splitting in magnetically sensitive lines, i.e.  lines with a high
Land\'e-g factor (e.g., Robinson, 1980; Marcy et al.  1989; Saar,
1996; Saar, 2001; Solanki, 1991 and references therein).  This is
usually achieved by comparing the profiles of magnetically sensitive
and insensitive absorption lines between observations and model
spectra. An alternate method that relies on the change in line
equivalent widths has been developed by Basri et al. (1992).  Modeling
the Zeeman effect on spectral lines in both cases requires the use of
a polarized radiative transfer code and knowledge of the Zeeman shift
for each Zeeman component in the magnetic field.  Furthermore, it
requires the observed lines to be isolated and that they can be
measured against a well-defined continuum. The latter becomes more and
more difficult in cooler stars since atomic lines vanish in the
low-excitation atmospheres and among the ubiquitous molecular lines
that appear in the spectra of cool stars. Measurements of stellar
magnetic fields carried out through detailed calculations of polarized
radiative transfer so far extend to stars as late as M4.5 (Johns-Krull
\& Valenti, 1996; 2000; Saar, 2001). In cooler objects, atomic lines
could not be used for the above-mentioned reasons, and because
suitable lines become increasingly rare.

Some magnetic field measurements in sun-like stars and early-M stars
are compiled in the articles by Saar (1996 and 2001), results are
available for spectral types G, K, and early M. Magnetic field
strengths and filling factors, i.e. the fraction of the star that is
filled with magnetic fields, apparently grow with later spectral type.
To my knowledge, no detection exists so far in stars of spectral type
F. At early-M spectral classes, filling factors in the (very active)
dMe stars approach unity with field strengths of several kilo-Gauss.

\begin{figure}[t]
  \epsfxsize=\textwidth \epsfbox{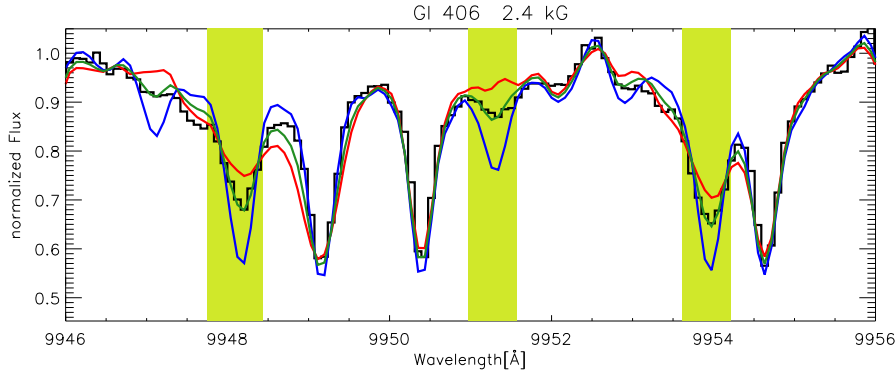}
  \caption{\label{fig:CNLeo}Measurements of magnetic flux in the dMe
    star Gl\,406 (CN~Leo). Scaled template spectra of a non-magnetic star 
(blue line) 
    and a magnetic star
(red line, $Bf \sim 3.9$\,kG)
    are shown. The fit to the data 
(green line) 
    is the interpolation between the two template spectra that best fits the
    data (see Reiners \& Basri, 2007). }
\end{figure}

\begin{figure}[t]
  \epsfxsize=\textwidth \epsfbox{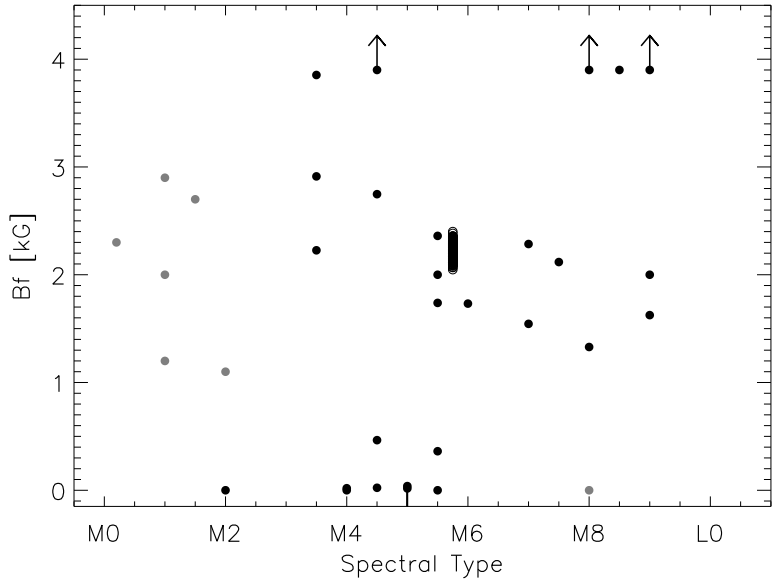}
  \caption{\label{fig:Mfields}Measurements of magnetic flux in M
    dwarfs. Grey circles are young objects, black circles are old
    objects. Uncertainties are usually on the order of several hundred
    Gauss.}
\end{figure}

In later stars and brown dwarfs, one alternative to atomic lines are
molecular bands with well separated individual lines so that they can
be distinguished from each other. Valenti et al. (2001) suggested that
FeH would be a useful molecular diagnostic for measuring magnetic
fields on ultra-cool dwarfs, but they point out that improved
laboratory or theoretical line data are required in order to model the
spectra directly. Reiners \& Basri (2006b) investigated the
possibility of detecting (and measuring) magnetic fields in FeH lines
of VLMs through comparison between the spectrum of a star with unknown
magnetic field strength and a spectrum of a star in which the magnetic
field strength is calibrated in atomic lines (early M dwarfs).
Although the Zeeman splitting in lines of molecular FeH is not
theoretically understood (but see Afram et al., 2007), the effect of
kilo-Gauss magnetic fields on the sensitive lines of FeH are easily
visible, and magnetic flux differences on the order of a kilo-Gauss
can be differentiated. This method was employed to measure the
magnetic flux in M-type objects down to spectral type M9 by Reiners \&
Basri (2007) and Reiners et al. (2007a, 2007b) discovering strong
magnetic fields in many ultra-cool dwarfs. An example of the detection
of a magnetic field in Gl\,406 (CN~Leo) is shown in
Fig.\,\ref{fig:CNLeo}. A list of measurements of magnetic flux in
M-type objects is compiled in Table\,\ref{tab:Mfields}.  Magnetic flux
$Bf$ is plotted as a function of spectral type in
Fig.\,\ref{fig:Mfields}.

\begin{table}
\centering
\begin{minipage}[t]{\columnwidth}
  \centering
  \caption{\label{tab:Mfields}Measurements of magnetic flux among M
    dwarfs. Young objects are shown in the lower part of the table.}
  \renewcommand{\footnoterule}{}
  \begin{tabular}{lcrrr}
    \noalign{\smallskip}
    \hline
    \hline
    \noalign{\smallskip}
    Name & Spectral Type & $v\,\sin{i}$  & $Bf$ & Ref\\
         &               &[km\,s$^{-1}$] & [kG]\\
    \noalign{\smallskip}
    \hline
    \noalign{\smallskip}   
    Gl~70       & M2.0 & $\le$3     &     $ 0.0$ & \footnote{Reiners \& Basri, 2007 (anchored at measurements from Johns-Krull \& Valenti, 2000)} \\
    Gl~729      & M3.5 &    4       &     $ 2.2$ & $^{a,}$\footnote{Johns-Krull \& Valenti, 2000} \\
    Gl~873      & M3.5 & $\le$3     &     $ 3.9$ & $^{a,b}$ \\
    AD~Leo      & M3.5 & $\approx$3 &     $ 2.9$ & $^{a,b}$ \\
    Gl~876      & M4.0 & $\le$3     &     $ 0.0$ & $^{a}$ \\
    GJ~1005A    & M4.0 & $\le$3     &     $ 0.0$ & $^{a}$ \\
    GJ~299      & M4.5 & $\le$3     &     $ 0.5$ & $^{a}$ \\
    GJ~1227     & M4.5 & $\le$3     &     $ 0.0$ & $^{a}$ \\
    GJ~1224     & M4.5 & $\le$3     &     $ 2.7$ & $^{a}$ \\
    YZ~Cmi      & M4.5 &    5       &     $>3.9$ & $^{a,b}$ \\
    Gl~905      & M5.0 & $\le$3     &     $ 0.0$ & $^{a}$ \\
    GJ~1057     & M5.0 & $\le$3     &     $ 0.0$ & $^{a}$ \\
    LHS~1070 A  & M5.5 &    8       &     $ 2.0$ & \footnote{Reiners et al., 2007b}\\
    GJ~1245B    & M5.5 &    7       &     $ 1.7$ & $^{a}$ \\
    GJ~1286     & M5.5 & $\le$3     &     $ 0.4$ & $^{a}$ \\
    GJ~1002     & M5.5 & $\le$3     &     $ 0.0$ & $^{a}$ \\
    Gl~406      & M5.5 &    3       &     2.1--2.4 & $^{a,}$\footnote{Reiners et al., 2007a} \\
    GJ~1111     & M6.0 &   13       &     $ 1.7$ & $^{a}$ \\
    VB~8        & M7.0 &    5       &     $ 2.3$ & $^{a}$ \\
    LHS~3003    & M7.0 &    6       &     $ 1.5$ & $^{a}$ \\
    LHS~2645    & M7.5 &    8       &     $ 2.1$ & $^{a}$ \\
    LP~412$-$31 & M8.0 &    9       &     $>3.9$ & $^{a}$ \\
    VB~10       & M8.0 &    6       &     $ 1.3$ & $^{a}$ \\
    LHS~1070 B  & M8.5 &   16       &     $ 4.0$ & $^{c}$ \\
    LHS~1070 C  & M9.0 &   16       &     $ 2.0$ & $^{c}$ \\
    LHS~2924    & M9.0 &   10       &     $ 1.6$ & $^{a}$ \\
    LHS~2065    & M9.0 &   12       &     $>3.9$ & $^{a}$ \\
    \noalign{\smallskip}
    \hline
    \noalign{\smallskip}
    CY Tau      & M1.0 &   11\footnote{Hartmann et al., 1986}&     $ 1.2$ & \footnote{Johns-Krull, 2007} \\
    DF Tau      & M1.0 &   19\footnote{Johns-Krull \& Valenti, 2001} &     $ 2.9$ & $^{f}$ \\
    DN Tau      & M1.0 &   10$^{g}$ &     $ 2.0$ & $^{f}$ \\
    DH Tau      & M1.5 &    8$^{g}$ &     $ 2.7$ & $^{f}$ \\
    DE Tau      & M2.0 &    7$^{g}$ &     $ 1.1$ & $^{f}$ \\
    AU Mic      & M2.0 &    8\footnote{Scholz et al., 2007}  &     $ 2.3$ & \footnote{Saar, 1994} \\
    2MASS1207   & M8.0 &   13       &     $<0.8$ & \footnote{Reiners \& Basri, submitted} \\
    \noalign{\smallskip}
    \hline
  \end{tabular}
\end{minipage}
\end{table}

The detections of strong magnetic flux in objects as late as spectral
type M9 show that magnetic field generation is very efficient in
completely convective objects, too. The idea of vanishing dynamo
action at the threshold to complete convection is certainly invalid
hence a lack of magnetic flux cannot be the reason for the weak
magnetic braking discussed above. Looking only at the currently
available measurements of magnetic flux, the opposite seems to be
true: Integrated flux grows with later spectral type, i.e. with deeper
convection zones (e.g. Saar, 1996). This impression, however, is
probably due to an observational bias. It is known that activity
(hence magnetic field generation) scales with rotation period or
Rossby number. At a given Rossby number (or period), hotter (and
larger) stars have higher surface rotation velocities than cooler
objects. The measurement of magnetic Zeeman splitting requires narrow
spectral lines in order to discriminate between Zeeman splitting and
other broadening mechanisms.  Thus, the strong fields that are
probably generated in more rapidly rotating, earlier stars cannot be
detected by current observational strategies.

Nevertheless, the result from the growing amount of magnetic flux
measurements in low mass stars is that a fully convective dynamo can
easily generate mean fields of kG-strength.  Reiners \& Basri (2007)
showed that H$\alpha$ activity in VLMs scales with magnetic flux just
as X-ray emission does in hotter stars (e.g. Saar, 2001).  From this
perspective, no change appears in the generation of magnetic activity
at the convection boundary.

Fully convective objects occupy a large region in the HR-diagram
(Fig.\,\ref{fig:HRdiagram}). Objects later than spectral type
$\sim$M3.5 are always completely convective, and more massive stars
develop a radiative core within the first $\sim$10\,Myr with the exact
time depending on their mass. Fully convective dynamos can also be
studied in pre-main sequence stars during their fully convective
phase. It is interesting to note that with one exception (the M2 dwarf
Gl~70), probably all objects shown in Fig.\,\ref{fig:Mfields} are
fully convective. Among the field stars, however, only the slowly
rotating ones have no strong (and large) magnetic field. The
rotation-activity connection seems to be intact at least in the sense
that rapid rotation involves the generation of a strong magnetic field
(Mohanty \& Basri, 2003; Reiners \& Basri, 2007). It is somewhat
disturbing that the only object later than M6 without a strong
magnetic field is the young accreting brown dwarf 2MASS
1207334--393254 (hereafter 2MASS~1207), which was found to drive a jet
(Whelan et al., 2007). The surface rotation velocity of 2MASS~1207 is
at the high end of velocities compared to its older (but more massive)
counterparts at similar spectral type ($v\,\sin{i}
\approx$\,5--12\,km\,s$^{-1}$), which is a good argument for a strong
magnetic field. The lack of a strong magnetic field on 2MASS~1207 may
indicate that magnetic field generation is also a function of age.
Reiners \& Basri (submitted) estimate that the magnetic field required
for magnetospheric accretion in 2MASS~1207 is only about 200\,G, i.e.
fully within the uncertainties of the magnetic flux measurement. They
speculate that during the accretion phase the magnetic field may be
governed by the accretion process rather than by the internal
generation through a convective dynamo, as is probably the case in
older objects.

The puzzle of magnetic field generation in fully convective stars is
currently receiving tremendous attention. Complementary strategies to
investigate the strength and topology of magnetic fields in fully
convective objects are applied very successfully. With the Zeeman
Doppler Imaging technique, Donati et al. (2006) showed that the
rapidly rotating fully convective M4 star V379~Peg exhibits a large
scale mostly axisymmetric magnetic topology. This finding is
apparently contradicting the idea that fully convective objects only
generate small scale fields. Browning (2007), performed simulations of
dynamo action in fully convective stars demonstrating that kG-strength
magnetic fields with a significant mean (axisymmetric) component can
be generated without the aid of a tachocline (see also Durney et al.,
1993; K\"uker \& R\"udiger, 1999; Chabrier \& K\"uker, 2006; Dobler et
al., 2006). From radio observations, Berger (2006) concluded that
fully convective objects must have strong magnetic fields, and he
finds that in very low mass stars the ratio of radio to X-ray emission
is larger than in sun-like stars. Hallinan et al. (2006) found
rotational modulation of radio emission from a rapidly rotating M9
dwarf. They conclude that the radio emission is difficult to reconcile
with incoherent gyrosynchrotron radiation, and that a more likely
source is coherent, electron maser emission from above the magnetic
poles. This suggestion, motivated by independent observations, also
requires the magnetic dipole (or multipole) to take the form of a
large-scale field with kG-strength. And recently, Berger et al. (2007)
showed in a multi-wavelength observation in an M9 object that X-ray,
H$\alpha$, and radio observations not necessarily correlate in time.
This raises the interesting question whether heating mechanisms differ
between sun-like stars and VLMs.

Another interesting aspect of magnetic fields in VLMs is the question
whether magnetism can effectively suppress convective heat transport.
Stassun et al. (2006, 2007) discovered the eclipsing binary brown
dwarf 2MASS J05352184--0546085 (2MASS~0535). They found that the more
massive primary surprisingly is cooler than the less massive
secondary. A possible explanation for this temperature reversal is a
strong magnetic field on the primary inhibiting convection (Chabrier
et al., 2007). Reiners et al. (2007c) discovered that H$\alpha$
emission in the primary is at least a factor of 7 stronger than in the
secondary. This supports the idea of a strong magnetic field on the
primary of 2MASS~0535. Effective cooling due to the presence of
magnetic fields on low-mass stars would have impact on the
mass-luminosity relation (see, e.g., Stauffer \& Hartmann, 1984;
Hawley et al., 1996; L\'opez-Morales, 2007).

\section{What happens at the threshold to complete convection?}
\label{sect:convective}

There is no doubt that around spectral type M3.5 a change in the
interior of main-sequence stars occurs, although the exact locus of
the convection boundary may shift towards later spectral types in the
presence of strong magnetic fields (Mullan \& McDonald, 2001). Stars
earlier than the convection boundary develop a radiative core and a
tachocline of shear at which a sun-like interface dynamo can work.
Stars cooler than the convection threshold do not harbor a tachocline
because no radiative core is developed. From the Sun we know that at
least the cyclic part of dynamo action is due to a dynamo operating at
the tachocline. This sort of dynamo can certainly not work in
completely convective objects, and if the interface dynamo was the
only one able to produce strong magnetic fields, objects later than
M3.5 simply could not produce magnetic flux and magnetic activity at
all.

Activity measurements across the convection boundary and the direct
detection of kG-strength magnetic fields in completely convective
objects rule out the possibility that the interface dynamo is the only
functioning type of dynamo in the stellar context. Obviously,
completely convective objects manage to generate large-scale fields
and probably also strong axisymmetric components. Furthermore, the
rapid rise of the fraction of objects exhibiting H$\alpha$ emission at
mid-M spectral class (West et al., 2004) could even be interpreted by
dynamo efficiency that is higher in the fully convective regime.
However, the clue to an understanding of dynamo activity in completely
convective objects is probably rotation and rotational braking. The
rising fraction of active stars at mid-M spectral type is possibly due
to the fact that fully convective stars in general are more rapidly
rotating than early-M dwarfs with radiative cores, and that the
rotation-activity relation is still working.

The rotational velocities of field M dwarfs show a remarkable rise
around spectral type M3.5 -- exactly where they are believed to become
fully convective. Braking timescales on either side of the convective
boundary differ by about an order of magnitude. The change in
rotational braking could be sufficient to explain the rising fraction
of active M dwarfs. Thus, the key question is: What is the reason for
the weak rotational braking in fully convective objects?

The strength of rotational braking depends on two main parameters: (I)
the strength of the magnetic field, and (II) the geometry of the
magnetic field. Because the strength of the magnetic field does not
seem to differ between partially and fully convective stars (the
latter may even have stronger fields), it is probably the magnetic
field geometry that differs between the two regimes. A sudden break in
magnetic field geometry at spectral type M3.5 could explain why no
rapid rotators are found at spectral types M0--M3, but are found at
later spectral types. A predominantly small-scale magnetic field would
lead to less braking than a large-scale field would. Indirect
observations of the magnetic field topology, however, suggest that
large scale axisymmetric fields do exist at least in some of them.
Theory also seems not to be in conflict with the generation of such
fields through a fully convective dynamo.  However, in order to decide
whether the magnetic topology differs on either side of the convection
boundary, we must convince ourselves that our observational methods
are sensitive to the aspects in question. In the case of fully
convective stars, this means that we must look for small scale fields,
which cannot be entirely excluded by the recent results of Zeeman
Doppler Imaging, because it is not sensitive to the very small scales.
One possibility is that a high density of closed loops near the
stellar surface prevents the outflow of a wind which normally would
brake the object.  Measurements of radio emission due to axisymmetric
fields also cannot exclude the presence of small scale fields on the
surface of fully convective objects.

Mass-dependent rotational braking can in principle explain the slope
of rotational velocities in L-dwarfs. Whether an abrupt change is
required in the magnetic properties of stars around the convection
boundary in order to explain the distribution of their rotation rates,
is not known. In order to answer this question, a larger,
statistically well defined sample is needed covering stars of
different ages. With the current instrumentation, this task can be
addressed within the near future.

\section{Summary}

During the last decade our picture of and beyond the bottom of the
main sequence has tremendously sharpened. In the ESO symposium on
``The Bottom of the Main Sequence -- and beyond'' in 1994, the last
spectral bin in an M-dwarf sample consisted of objects of spectral
class M5.5+ (Hawley, 1995). Since then, our view of VLMs rapidly
reached out much further thanks to several surveys and large
telescopes capable of collecting high quality data of faint stars and
brown dwarfs. We have followed the main sequence down to the faintest
stars and beyond it to brown dwarfs cooler than 1000\,K. In
particular, we are now scrutinizing the two regions of major changes
in the physics of low mass objects; the threshold to complete
convection and the threshold to brown dwarfs. From an observational
point, there is still not much to say about the latter; it does not
seem to have very large impact on the surface properties of low mass
objects (and its activity, Mokler \& Stelzer, 2002).  Discriminating
between low mass stars and heavy brown dwarfs is still a challenge.

The threshold to complete convection is expected to influence the
observable properties of low mass stars. Observations at either side
of the convection boundary can be sorted into two groups, those
without evidence for a change in magnetic field generation and those
with evidence for it:

\subsubsection*{Parameters not affected by the threshold to complete
  convection:}
\begin{enumerate}
\item The mean levels of activity measurements in X-ray, H$\alpha$,
  and other tracers do not provide significant evidence for a change
  in magnetic field generation. In particular, mean chromospheric and
  coronal emission is not observed to diminish around spectral type
  M3.5. At later spectral class (around M9), activity decreases with
  lower temperature. The reason for this is probably not less dynamo
  efficiency but the growing neutrality of the cooler atmosphere.
\item Direct measurements of magnetic fields show that kG-strength
  magnetic fields are still generated in fully convective stars. Such
  fields are detected in objects as late as M9, which is far beyond
  the convection boundary. Fully convective stars apparently harbor
  strong fields that occupy a large fraction of their surface.
\item Zeeman Doppler Imaging and measurements of persistent periodic
  radio emission suggest that fully convective objects still harbor
  large scale axisymmetric magnetic field components.
\end{enumerate}

\subsubsection*{Parameters indicating a break at the convection
  boundary:}
\begin{enumerate}
\item Around spectral type M3.5, a sudden break is observed in the
  distribution of rotational velocities $v\,\sin{i}$. On its cool
  side, hardly a single field M dwarf is known with detected
  rotational broadening; members of the young and the old disk
  populations are rotating at a very low rate. On the cool side of the
  boundary, however, many young disk objects are known with detected
  rotation velocities, some of them rotating at several ten
  km\,s$^{-1}$. At the threshold to complete convection, the braking
  timescale changes by about an order of magnitude.
\item The highest normalized emission found in X-ray or H$\alpha$
  emission of field dwarfs is about an order of magnitude larger at
  spectral type M4 than it is around M2.
\item The fraction of active stars exhibits a steep rise from below
  10\,\% at spectral class M3 to more than 50\,\% at M5. 
\end{enumerate}

The two last points in favor of a change at the convection boundary
may be explained by the higher fraction of rapid rotators among
completely convective objects (point 1). A possible scenario is that
in fully convective objects weaker braking leads to higher rotation
velocities in objects observed in the field. With the rotation
activity connection still working, this implies a higher fraction of
active stars. Higher rotation velocities also lead to higher maximum
emission levels. If this scenario is true, the key question is: ``Why
is rotational braking less efficient in fully convective objects?''
One possible answer is a change in the magnetic topology due to a
different dynamo process, and this possibility is currently receiving
high attention.

Related open questions are the following:

\begin{itemize}
\item In stars with radiative cores, what is the fraction of the
  magnetic flux that is generated within the convection zone, and what
  is the fraction generated in the interface dynamo?
\item Does (and to what extent does) the rotation activity connection
  hold in fully convective objects?
\item Does the fully convective dynamo depend on differential rotation
  (and how does the interface dynamo)?
\item What is the magnetic topology of fully convective objects (and
  what is the topology of rapidly rotating sun-like stars)?
\item How does magnetic field generation evolve with age?
\item Are strong magnetic fields generated in brown dwarfs, too?
\end{itemize}

The investigation of the properties of VLMs and of stars around the
convection boundary has long been hampered by technical difficulties
obtaining suitable data (and a sufficient amount of them). Today, our
observational equipment is certainly suited to tackle some of these
questions. Theoretical considerations together with numerical
calculations are pushing and questioning our understanding of low mass
star and brown dwarf physics. Within the near future, we can certainly
expect answers to some of the open questions at the bottom of the main
sequence and beyond it.

\subsection*{Acknowledgements}

I thank the Astronomische Gesellschaft for the \emph{Ludwig-Biermann
  Prize} 2007, and I acknowledge research funding from the DFG through
an Emmy Noether Fellowship (RE 1664/4-1).

\subsection*{References}

{\small

  \bref Afram, N., Berdyugina, S.V., Fluri, D.M., Semel, M., Bianda,
  M., Ramelli, R., 2007, A\&A, 473, L1

  \bref Baraffe, I., Chabrier, G., Allard, F., \& Hauschildt, P.H., 1998,
  A\&A, 337, 403

  \bref Baraffe, I., Chabrier, G., Allard, F., \& Hauschildt, P.H., 2002,
  A\&A, 382, 563
  
  \bref Barnes, S.A., 2007, \texttt{arXiv:0704.3068}

  \bref Basri, G., 2000, ARA\&A, 38, 485

  \bref Basri, G., Marcy, G.W., \& Valenti, J.A., 1992, ApJ, 390, 622

  \bref Benedict, F., et al., 1998, AJ, 116, 429

  \bref Berger, E., 2006, ApJ, 648, 629

  \bref Berger, E., 2007, \texttt{arXiv:0708.1511}

  \bref Bessell, M.S., 1991, AAS 83, 357 

  \bref Browning, M.K., 2007, ApJ in press, \texttt{arXiv:0712.1603}

  \bref Burgasser, A.J.Liebert, J., Kirkpatrick, J.D., Gizis, J.E.,
  2002, AJ, 123, 2744

  \bref Cayrel de Strobel G., Soubiran C., Friel E.D., Ralite N.,
  Francois P., 1997, A\&AS, 124, 299

  \bref Chabrier, G., Baraffe, I., \& Plez, B., 1996, ApJ, 459, L91

  \bref Chabrier, G., \& Baraffe, I., 2000, ARAA, 38, 337
  
  \bref Chabrier, G., Baraffe, I., Allard, F., \& Hauschildt, P.H.,
  2005, ASP Conf. Ser., \texttt{arXiv:astro-ph/0509798}

  \bref Chabrier, G., \& K\"uker, M., 2006, A\&A, 446, 1027

  \bref Chabrier, G., Gallardo, J., \& Baraffe, I., 2007, A\&A, 472,
  L17

  \bref Chaboyer, B., Demarque, P., \& Pinsonneault, M.H., 1995, ApJ,
  441, 876

  \bref Clemens, J.C., Reid, I.N., Gizis, J.E., \& O'Brien, M.S.,
  1998, ApJ, 496, 352
  
  \bref Copeland, H., Jensen, J.O, \& Jorgensen, H.E., 1970, A\&A, 5,
  12

  \bref Dahn, C.C., et al., 2002, AJ, 124, 1170 
  
  \bref Dobler, W., \& Stix, M., \& Brandenburg, A., 2006, ApJ, 638,
  336

  \bref Donati, J.-F., Forveille, T., Collier Cameron, A., Barnes,
  J.R., Delfosse, X., Jardina, M.M., \& Valenti, J.A., 2006, Science,
  311, 633

  \bref D'Antona, F., \& Mazzitelli, I., 1996, ApJ, 456, 329

  \bref Delfosse, X., Forveille, T., Perrier, C., \& Mayor, M., 1998,
  A\&A, 331, 581

  \bref Durney, B.R., \& Latour, J., 1978, Geophys. Astrophys. Fluid
  Dynamics, 1978, 9, 241

  \bref Durney, B.R., De Young, B.S., \& Roxburgh, I.W., 1993,
  Sol.Phys.,145, 207

  \bref ESA, 1997, The Hipparcos \& Tycho Catalogues, ESA-SP 1200

  \bref Fleming, T.A., Giampapa, M.S., \& Schmitt, J.H.M.M., 2000,
  ApJ, 533, 372
  
  \bref Golimowski, D.A., Leggett, S.K., Marley, M.S., et al., 2004,
  AJ, 127, 3516

  \bref Hallinan, G., Antonova, A., Doyle, J.G., Bourke, S., Brisken,
  W.F., \& Golden, A., 2006, ApJ, 653, 690
  
  \bref Hartmann, L., Hewett, R., Stahler, S., Mathieu, R.D., 1986, ApJ, 309, 275

  \bref Hawley S.L., Gizis J.E., Reid I.N., AJ, 1996, 112, 2799

  \bref Hauck, B. \& Mermilliod, M., 1998, A\&AS, 129, 431
  
  \bref Herbst, W., Eisl\"offel, J., Mundt, R., \& Scholz, A.,
  \texttt{arXiv:0603673}
  
  \bref Johns-Krull, C., \& Valenti, J. A. 1996, ApJ, 459, L95

  \bref Johns-Krull, C., \& Valenti, J.A., 2000, ASPC, 198, 371

  \bref Johns-Krull, C., \& Valenti, J. A. 2001, ApJ, 561, 1060

  \bref Johns-Krull, C., 2007, ApJ, 664, 975

  \bref Kawaler, S.D., 1988, ApJ, 333, 236

  \bref Kiraga, M., \& Stepie\'n, 2007, \texttt{arXiv:0707.2577}

  \bref Kirkpatrick, J.D., 2005, ARA\&A, 43, 195

  \bref Krishnamurti, A., Pinsonneault, M.H., Barnes, S., \& Sofia,
  S., 1997, ApJ, 480, 303

  \bref K\"uker, M., \& R\"udiger, G., 1999, A\&A, 346, 922

  \bref Leggett, S.K., 1992, ApJS 82, 351

  \bref Liebert, J., Lirkpatrick, J.D., Cruz, K.L., Reid, I.N.,
  Burgasser, A., Tinney, C.G., \& Gizis, J.E., 2003, AJ, 125,343

  \bref Liebert, J., \& Probst, R.G., 1987, ARAA, 25, 473

  \bref L\'opez-Morales, M., 2007, ApJ, 660, 732

  \bref Marcy, G.W., \& Basri, G., 1989, ApJ, 345, 480

  \bref Mestel, 1984, in Third Cambridge Workshop on Cool Stars,
  Stellar Systems, and the Sun, ed. S.L.  Baliunas \& L.  Hartmann
  (New York: Springer), 49

  \bref Meyer, F., \& Meyer-Hofmeister, E., 1999, A\&A, 341, L23

  \bref Mohanty, S., Basri, G., Shu, F., Allard, F., Chabrier, G.,
  2002, ApJ, 571, 469

  \bref Mohanty, S., \& Basri, G., 2003, ApJ, 583, 451

  \bref Mohanty, S., Jayawardhana, R., \& Basri, G., 2005, ApJ, 626,
  498
 
  \bref Mokler, F., \& Stelzer, B., 2002, A\&A, 391, 1025

  \bref Noyes, R.W., Hartmann, L.W., Baliunas, S.L., Duncan, D.K., \&
  Vaughan, A.H., 1984, ApJ, 279, 763

  \bref Ossendrijver, M., 2003, A\&AR, 11, 287

  \bref Patten, B.M., \& Simon, T., 1996, ApJSS, 106, 489

  \bref Pettersen, B.R., 1983, in Byre P.B., Rodono M., eds., Activity
  in Red-dwarf Stars, Reidel, Dordrecht, p.17

  \bref Pizzolato, N., Maggio, A., Micela, G., Sciortino, S., \&
  Ventura, P., 2003, A\&A, 397, 147

  \bref Randich, S., 1998, ASP Conf. Ser., 154, 501

  \bref Reid I.N., Hawley S.L., Gizis J.E., AJ, 1995, 110, 1838

  \bref Reiners, A., 2007, A\&A, 467, 259

  \bref Reiners, A., \& Basri, G., 2006a, AJ, 131, 1806

  \bref Reiners, A., \& Basri, G., 2006b, ApJ, 644, 497

  \bref Reiners, A., \& Basri, G., 2007, ApJ, 656, 1121

  \bref Reiners, A., Schmitt, J.H.M.M., \& Liefke, C., 2007a, A\&A
  466, L13

  \bref Reiners, A., Seifahrt, A., Siebenmorgen, R., K\"aufl, H.U., \&
  Smette, Al., 2007b, A\&A, 471, L5

  \bref Reiners, A., Seifahrt, A., Stassun, K.G., Melo, C., \&
  Mathieu, R.D., 2007c, ApJL, in press, \texttt{arXiv:0711.0536}

  \bref Robinson, R.D., 1980, ApJ, 239, 961

  \bref Robrade, J., \& Schmitt, J.H.M.M., 2005, A\&A, 435, 1077

  \bref Saar, S.H., 1994, IAU Symp. 154, Infrared Solar Physics, eds.
  D.M. Rabin et al., Kluwer, 493

  \bref Saar, S.H., 1996, in IAU Symp. 176, Stellar Surface Structure,
  eds., Strassmeier, K., J.L. Linsky, Kluwer, 237

  \bref Saar, S.H., 2001, ASPCS 223, 292
  
  \bref Scholz, A., \& Eisl\"offel, J., 2004, A\&A, 421, 259
  
  \bref Scholz, A., \& Eisl\"offel, J., 2007, MNRAS, 381, 1638
  
  \bref Scholz, A., Coffey, J., Brandeker, A., \& Jayawardhana, R., 2007, ApJ, 662, 1254

  \bref Schmidt, S.J., Cruz, K.L., Bongiorno, B.J., Liebert, J., \&
  Reid, I.N., 2007, AJ, 133,2258

  \bref Schmitt, J.H.M.M., 1995, ApJ, 450, 392

  \bref Siess, L., Dufour, E., \& Forestini, M., 2000, A\&A, 358, 593

  \bref Simon, T., 2001, ASP Conf. Ser., 223, 235

  \bref Solanki, S.K., 1991, RvMA, 4, 208

  \bref Stauffer, J.R., \& Hartmann, L.W., 1986, ApJS, 61, 531

  \bref Stassun, K.G., Mathieu, R.D., \& Valenti, J.A., 2006, Nature,
  440, 311

  \bref Stassun, K.G., Mathieu, R.D., \& Valenti, J.A., 2007, ApJ,
  664, 1154

  \bref Sterzik, M.F., \& Schmitt, J.H.M.M, 1997, AJ, 114, 167

  \bref Takalo, L.O., \& Nousek, J.A., 1988, ApJ, 326, 779

  \bref Taylor, B.J., 1995, PASP, 107, 734

  \bref Torres, C.A.O., Busko, I.C., \& Quast, G.R., 1983, in Byre
  P.B., Rodono M., eds., Activity in Red-dwarf Stars, Reidel,
  Dordrecht, p.175

  \bref Valenti, J.A., Johns-Krull, C.M., \& Piskunov, N.P., 2001, ASP
  223, 1579

  \bref Voges, W., et al., 1999, A\&A, 349, 389

  \bref West, A.A., et al., 2004, AJ, 128, 426

  \bref Whelan, E.T., Ray, T.P, Randich, S., Bacciotti, F.,
  Jayawardhana, R., Testi, L., Natta, A., \& Mohanty, S., 2007, ApJ,
  659, L45

  \bref Zapatero Osorio, M.R., Mart\'in, E.L., Bouy, H., Tata, R.,
  Deshpande, R., \& Wainscoat, R.J., 2006, ApJ, 647, 1405

}

\vfill

\end{document}